\documentclass[10pt,twocolumn,letterpaper]{article}

\usepackage{cvpr}
\usepackage{times}
\usepackage{epsfig}
\usepackage{graphicx}
\usepackage{amsmath}
\usepackage{amssymb}
\usepackage[breaklinks=true,bookmarks=false]{hyperref}

\cvprfinalcopy % *** Uncomment this line for the final submission

\def\cvprPaperID{****} % *** Enter the CVPR Paper ID here

\begin{document}

%%%%%%%%% TITLE
\title{Image Compression with Recurrent Neural Network and Generalized Divisive Normalization}

\author{Khawar Islam\textsuperscript{1}, L. Minh Dang\textsuperscript{1}, Sujin Lee\textsuperscript{2}, Hyeonjoon Moon\textsuperscript{1,2} \\
Computer Vision and Pattern Recognition Lab\textsuperscript{1}, Sejong University, South Korea\\
Department of Artificial Intelligence\textsuperscript{1,2}, Sejong University, South Korea \\
{\tt\small \{khawar,minhdl\}@sju.ac.kr, genegraphy@gmail.com, hmoon@sejong.ac.kr}}

\maketitle
%\thispagestyle{empty}

%%%%%%%%% ABSTRACT
\begin{abstract}
   Image compression is a method to remove spatial redundancy between adjacent pixels and reconstruct a high-quality image. In the past few years, deep learning has gained huge attention from the research community and produced promising image reconstruction results. Therefore, recent methods focused on developing deeper and more complex networks, which significantly increased network complexity. In this paper, two effective novel blocks are developed: analysis and synthesis block that employs the convolution layer and Generalized Divisive Normalization (GDN) in the variable-rate encoder and decoder side. Our network utilizes a pixel RNN approach for quantization. Furthermore, to improve the whole network, we encode a residual image using LSTM cells to reduce unnecessary information. Experimental results demonstrated that the proposed variable-rate framework with novel blocks outperforms existing methods and standard image codecs, such as George's ~\cite{002} and JPEG in terms of image similarity. The project page along with code and models are available at \href{https://github.com/khawar512/cvpr_clic_image_compression.git}{\color{red}{https://github.com/khawar512/cvpr\_image\_compress}}
\end{abstract}

%%%%%%%%% BODY TEXT
\section{Introduction}

Recent deep learning approaches for lossy image compression have been gained significant interest in machine learning and achieves more promising results. It plays a significantly important role in streaming a large amount of image and video data under adequate storage, and low bandwidth \cite{013,012,010}. Deep learning has made tremendous contributions in neural network-based image coding. \par
Various image compression methods ~\cite{006,008,002} have achieved remarkable performance. Ballé et al. \cite{004} proposed a generalized divisive normalization (GDN) based framework which comprises three main steps: convolutional layer, sub-sampling scheme, and nonlinear GDN layers. Recently, a novel model for learned image compression is context-adaptive \cite{005} receives significant performance and outperforms all image codec. Johannes et al. \cite{004} designed a simple hyperprior model to add a greater number of bits in the entropy module. Minnen et al. \cite{011} combined an auto-regressive with a hierarchical module to obtain better results in the context of image compression. This work \cite{005} constructed a similar technique \cite{004} by combining two different context models; one is a bit consuming, and the other one is bit-free contexts to understand a context-adaptive model. \par
Although CNN-based image compression methods \cite{005,010,011}, already improved compression accuracy, commonly mostly methods are increasing more layers and make deeper network to achieve compression results. These methods substantially increased model complexity and computational cost. As far as we know, the development of image compression networks based on RNN is relatively small compared to CNN and auto-encoders. These techniques ~\cite{002,007} proposed a full image resolution network using residual scaling, recurrent neural network (RNN), and entropy coding based on deep learning. This network simultaneously generated three models during training. \par
In this paper, two spatial adaptive blocks called analysis and synthesis are proposed in the variable-rate encoder and decoder side based on a convolution layer and generalized divisive normalization process. Then, it embeds with a variable-rate framework. The analysis block generates a powerful spatial and channel representation with a down-sampling feature for encoding. Simultaneously, synthesis block up-sampling the decoded image. To the best of our knowledge, constructing recurrent neural network-based image compression networks is very limited. This is the first work that employs a generalized divisive normalization scheme in a variable rate network to handle multiple bit rates in the RNN network. Furthermore, visual quality performance and scalability of the proposed blocks and end-to-end network are jointly validated by the popular evaluation with PSNR and MS-SSIM. Comprehensive experiments have been performed on Kodak to describe that the proposed blocks and network jointly achieve significant performance by comparing standard image codecs JPEG, BPG, WebP, and recent method George's ~\cite{002}. 

\begin{figure}[t]
    \centering
    \includegraphics[width=\linewidth]{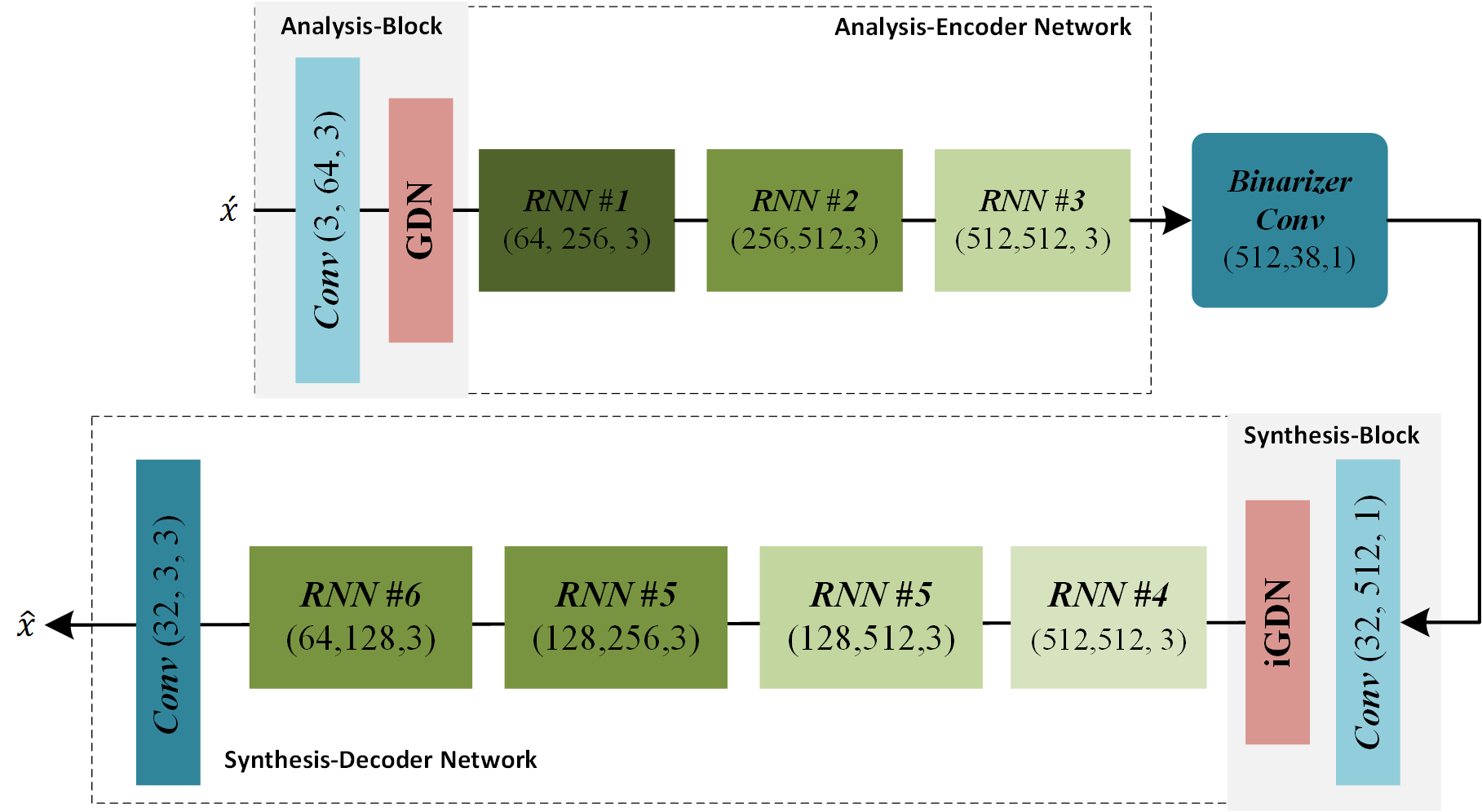}
    \caption{Iterative architecture of image compression framework based on recurrent neural network. Each input patch was first passed to the analysis-encoder block to enrich image representation. Similarly, the synthesis-decoder block reconstructs a decoded image with the help of recurrent neural network cells.}
     \label{fig:Fig_1}
\end{figure}

\section{Proposed Network}
The overall architecture of image compression with two blocks is presented in Fig \ref{fig:Fig_1}. Motivated by the recent development of image compression networks ~\cite{002,003}. There are three modules with two additional novel blocks in the end-to-end framework, i.e., encoder network, analysis block, binarizer, decoder network, and synthesis block. Image patches are directly given to the analysis block as an input that generates latent features using the proposed analysis-encoder block. Then, the latent representation will be passed to RNN cells for sequence generation. Further, binarizer quantized latent representation using a binary RNN approach and send it to the decoder network and proposed synthesis block with decoder network to construct the final image. Binarization scheme is utilized as ~\cite{002}. The image compression framework aims to reconstruct the high-quality image at a given bitrate. It is an important part of developing a better entropy model. Meanwhile, the process defined in ~\cite{002}, and construct encoder network which contains one convolutional layer, analysis-encoder block, and three RNN cells, stateless binarizer contains linear convolutional layer and decoder contains single convolutional layers with synthesis decoder block with four RNN cells and one more deconvolutional layer. The single iteration of the end-to-end framework is represented in Eq 1. The entire framework architecture is presented in Fig \ref{fig:Fig_1}. 
\begin{equation}
\begin{array}{l}
b_{t}=\operatorname{Bin}\left(\operatorname{Enc}_{t}\left(r_{t}-1\right)\right) \\
\hat{x}_{t}=\operatorname{Dec}_{t}\left(\operatorname{bin}_{t}\right)+\gamma \hat{x}_{t-1}
\end{array}
\end{equation}
\begin{equation}
r_{t}=x-\hat{x}_{t}, \quad r_{0}=x, \quad \hat{x}_{0}=0
\end{equation}
Where $Enc$ and $Dec$ are the encoder and decoder with iteration $t$, \emph{bin\textsubscript{t}} is binary representation \emph{\^{x}\textsubscript{t}} is the reconstruction version of actual image $x$ and $\gamma$ = 0  is one shot reconstruction. The spatial adaptive blocks and end-to-end variable-rate image compression architecture will be presented in next two sections. The training process of image compression network is optimized by adopting the loss at each iteration based on actual weighted and predicted value. 
\begin{equation}
L_{1}=\sum_{i=1}^{n}\left|y_{\text {true}}-y_{\text {predicted}}\right|
\end{equation}
The overall loss at each iteration of the variable-rate framework is:
\begin{equation}
L_{1}=\beta \sum_{t}\left|r_{t}\right|
\end{equation}

\subsection{Analysis Block}
An effective spatial adaptive analysis block is proposed based on spatial information. The proposed block is shown in Fig \ref{fig:Fig_1}. The analysis block combines generalized divisive normalization scheme ~\cite{001} and utilized as a nonlinear transformation between convolution layers in ~\cite{005}, and ~\cite{004}. This parametrizes the nonlinear transformation block, which is more suitable with gaussian data and natural images. Block consists of a uniform scalar quantization scheme; it constructs a parameterized form of feature vector from the latent representation of the input image vector space. In ~\cite{004}, a generalized divisive normalization scheme captures spatial information in latent representation is used in the block for nonlinear transformation. This technique is never utilized in recurrent neural networks. Some work ~\cite{010} used a separate attention module on the encoder side and a gaussian mixture model and decoder enhancement module on the decoder side to reconstruct a better-decoded image. The structure of our analysis-encoder block contains three input channels and $64$ out channels. The kernel size of layers is $3$, and stride is $1$. The analysis block consists of the convolution layer, generalized divisive normalization technique. Every step begins with affine convolution.
\begin{equation}
v_{i}^{(k)}(m, n)=\sum_{j}\left(h_{k, i j} * u_{j}^{(k)}\right)(m, n)+c_{k, i}
\end{equation}
in which $i\textsuperscript{th}$  is input channel of $k\textsuperscript{th}$ step at spatial location $(m,n)$ as $v\textsubscript{i}\textsuperscript{(k)} (m, n).$
$\emph{x}$ is input image vector corresponds to $u\textsubscript{i}\textsuperscript{(0)} (m, n)$, and output vector $\emph{y}$ is $u\textsubscript{i}\textsuperscript{(3)}(m, n)$. $\ast$ represents 2D convolution:
\begin{equation}
w_{i}^{(k)}(m, n)=v_{i}^{(k)}\left(s_{k} m, s_{k} n\right)
\end{equation}
Where downsampling factor at step $k$ represented with $s\textsubscript{k}$. Every step is followed by the GDN operation. Now, Equ (7), defined all two stages of GDN operation where $\beta\textsubscript{k,i}$ and $\gamma\textsubscript{k,ij}$ are two scale and bias parameters of normalization operation.
\begin{equation}
u_{i}^{(k+1)}(m, n)=\frac{w_{i}^{(k)}(m, n)}{\left(\beta_{k, i}+\sum_{i} \gamma_{k, i j}\left(w_{j}^{(k)}(m, n)\right)^{2}\right)^{\frac{1}{2}}}
\end{equation}
\subsection{Synthesis Block}
Considering the problem of the final reconstructed image, the image may contain several artifacts. An effective synthesis block is proposed at the decoder side after quantization to keep the decoded image's information and quality. As illustrated in Fig \ref{fig:Fig_1}, several convolutional and inverse GDN layers to reconstruct the input image. Same as the aforementioned analysis block, the decoder module is comprised of two phases in which all the processes are inversed. After iGDN operation, the feature vector passes into RNN cells for further reconstruction, and the deconvolutional layer forms the final reconstructed image. Every step starts from the convolutional layer, and then inverse GDN operation calculates as follows:
\begin{equation}
\begin{aligned}
\widehat{w}_{i}^{(k)}(m, n) &=\hat{u}_{i}^{(k)}(m, n) \cdot\left(\hat{\beta}_{k, i}\right.\\
&\left.+\sum_{j} \hat{\gamma}_{k, i j}\left(\hat{u}_{j}^{(k)}(m, n)\right)^{2}\right)^{\frac{1}{2}}
\end{aligned}
\end{equation}
Inverse GDN comprises three steps, with inverse operation at each step, downsampling, and Inverse GDN. $\hat{u}_{i}^{(k)}(m, n)$ is the input $k\textsuperscript{th}$ synthesis step. The design of our synthesis block contains $38$ input channels and $512$ output channels. The kernel size of layers is $1$ and stride is $1$ with padding $0.$
\subsection{RNN Cell}
\begin{figure}
    \centering
    \includegraphics[scale=0.35]{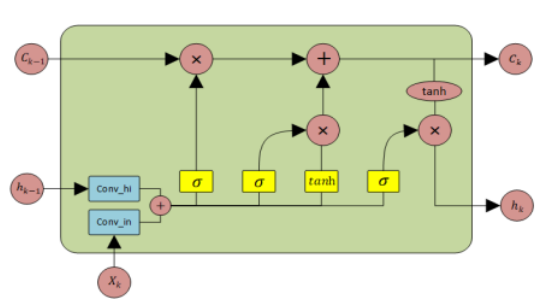}
    \caption{Overview of recurrent neural network cell.}
     \label{fig:Fig_2}
\end{figure}

The image compression network comprises RNN units and a pixel depth scheme. In each iteration, RNN units are utilized for the extraction of features from the image. Continuously, memorize the residual state in each iteration process and reconstruct the image. The entire network uses seven RNN cells, three cells are encoder side, and four units are on the decoder side. In every iteration, the estimated outcome
of the previous cell passes to the hidden layer for further iteration. The simplified view of the RNN cell is illustrated in Fig \ref{fig:Fig_2}. The hidden and memory layer state of RNN cell is $c_{k-1}$ and $x_{k-1}$. The input feature vector for $k\textsuperscript{th}$ iteration is $x_k$ that is equal to the upper layer output in this iteration. Each cell comprises of two convolution layers; $conv\_h\textsubscript{i}$ works for input feature vector $x_k$, and $conv\_h\textsubscript{i}$ works for hidden layer.

In the above aforementioned figure, $x_k$, $c_{k-1}$ and $x_{k-1}$ are input vectors. Where $X$ denotes element-wise multiplication. Similarly, + represents element-wise
addition $c_k$, $h_k$ are the output vectors for this cell and input vectors for the next cell. Here, $x_k$, $c\textsubscript{k-1}$ and $x\textsubscript{k-1}$ are the input vectors from convolutional layer. $c_k$ and $h_k$ are the calculation methods in each iteration are defined as:
\begin{equation}
\begin{array}{l}
{[f, i, \widetilde{C}, o]^{T}=[\sigma, \sigma, \tanh , \sigma]^{T}} \\
\left(\operatorname{conv}_{-} i n\left(x_{k}\right)+\operatorname{con} v_{-} h i\left(h_{k-1}\right)\right)
\end{array}
\end{equation}

\begin{equation}
c_{k}=f \odot c_{k-1}+i \odot \tilde{C}
\end{equation}
\begin{equation}
h_{k}=o \odot \tanh \left(c_{k}\right)
\end{equation}

\subsection{Entropy Coding}
We enhance the performance of the network by using different hidden values in binarizer module. As shown in Fig \ref{fig:Fig_3}. The approach of pixel RNN \cite{008} as a binary RNN for image compression with a single convolution layer. The quantization approach describes in \cite{003} and applies quantization noise during training. The binarizer part comprises a single linear convolutional layer with an activation function. The entire binarizer module generates binary codes within an interval of $(1,-1)$ with the help of the sign function. The input feature vectors $H$x$W$x$3$ could be compressed in $(H/16)$ x $(W/16)$ x $38$ binary code. It means that achieving bit per pixel in every iteration, which is $1/8$ and the compression ratio of the image is $i\textsuperscript{th}$ iteration is $i/192$. The binarizer layer takes $512$ as input channels, and $32$ is the output channels with kernel size $1$.

\section{Experiments}
In order to perform the comparison of proposed work with recent standard image compression codecs and existing deep learning-based frameworks are discussed.
\subsection{Implementation Details}
Different experiments are performed on large-scale datasets. In our experiment, $20745$ high-quality images from Flickr are adopted and created a subset dataset that contains $3600$ training images and generates random patches of $32$×$32$ for training. The total number of patches is $960,8744$ with batch size $16$ and on each epoch $(960,8744/16)$ is $600,547$ iterations. All networks included encoder, binarizer, and decoder were trained using PyTorch, with Adam optimizer. We trained the network for a $10$ epoch using a batch size of $16$. The entire framework was trained using learning rates $0.0005$ for ten epochs. Each tensor of the image represented four values. The tensor contained the batch size $16$, image height and width are $32$, and $C$ represented the number of channels in the color image, $3$.
\begin{figure}[t]
    \centering
    \includegraphics[scale=0.5]{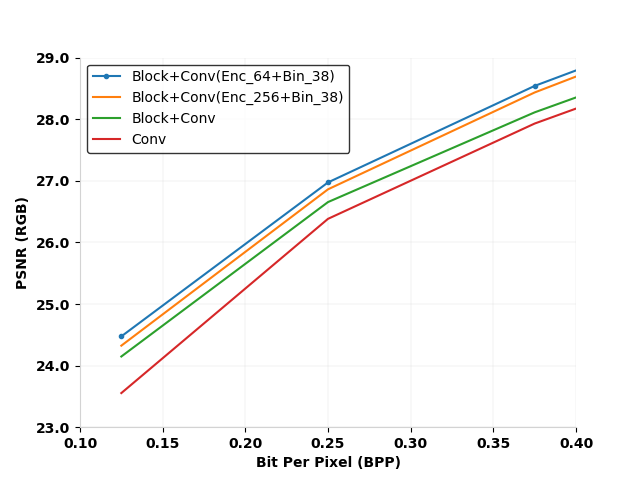}
    \caption{The comparison of rate-distortion curves with different hidden values.}
     \label{fig:Fig_3}
\end{figure} 
\begin{figure*}
    \centering
    \includegraphics[scale=0.7]{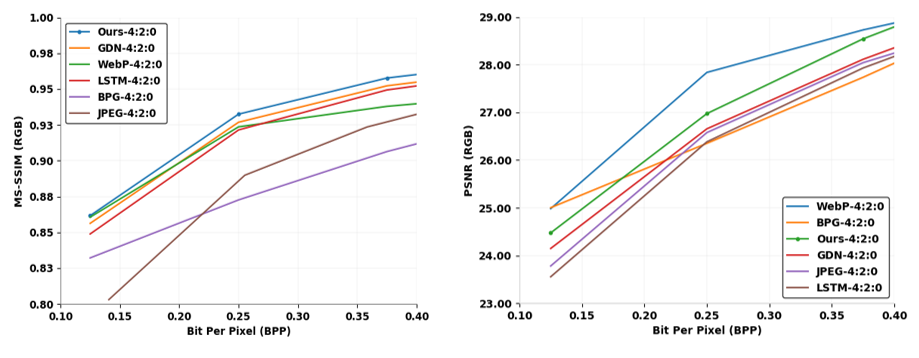}
    \caption{Comparative results of variable-rate network with analysis and synthesis block with recent approaches on kodak benchmark in terms of MS-SSIM (left side) and PSNR (right side) vs. BPP.}
     \label{fig:Fig_4}
\end{figure*}

\subsection{Discussion on Architecture Improvement}
We conduct several experiments to improve George's ~\cite{002} architecture. We have found that the output channels in our architecture do not give much improvement. Firstly, we add GDN block in encoder and decoder modules, which substantially improves PSNR and SSIM. Secondly, we modify the hidden values in binarizer as $[512, 32]$ to $[512, 38]$, and in decoder hidden values as $[32, 512]$ to $[38, 512]$. Thirdly, we changed the encoder hidden values as $[64,128]$ to $[256, 512]$ combining experiments with GDN. Lastly, we set the encoder values as $[64, 128]$ and remain changed in binarizer module, correspondingly. This work employed RNN cells with the implementation of GDN with the hyperparameters. The findings of our experiments are presented in Fig \ref{fig:Fig_3}. Experiments have shown that the hidden values in binarizer module with GDN blocks $[512, 38]$ could be better than hidden values $[512, 32]$ in terms of PSNR. This improvement remains the same in MS-SSIM. Finally, in the above-aforementioned experiments, the value of $Block+Conv(Enc64+Bin38)$ has been utilized in our network. \par

\textbf{GDN/iGDN vs Conv.} To evaluate the visual quality of the analysis and synthesis blocks. We integrate them into our network. In this model, both blocks are removed from George's ~\cite{002} architecture. We followed a straightforward propagation in the network. As shown in Fig \ref{fig:Fig_3}. the only Conv layer does not perform well. Then, we added a convolution and GDN layer in all forward propagation operations. As presented in Fig \ref{fig:Fig_3}, the architecture with analysis and synthesis block contains significantly better than George's ~\cite{002} in terms of visual similarity and rate-distortion.

\textbf{Coding efficiency.} To compare the performance of standard image codecs and existing methods, experiment on a widely used Kodak dataset ($24$ testing images) has
been conducted. The rate-distortion curves by taking the different quality factors of each Kodak image are shown in Fig \ref{fig:Fig_4}. In each image, $20$ images are generated with different quality starting from lower value to higher value, i.e., $Q = {5, 10, 15, 20....100}$. The residual image for the BPG codec is encoding with different quantizer parameters starting from $QP = {51, 46, 41...}$, and chroma sampling (YUV4:2:0 format). We evaluate our image compression network's visual quality and existing variable rate methods George's ~\cite{002}, and standard image codecs JPEG, HEVC intra-coding based BPG, and WebP. Perceptual full-reference quality metrics for comparing the original image and reconstructed image are employed. Thus, MS-SSIM and PSNR well-defined quality metric designed for image compression algorithms are used. MS-SSIM on RGB images is applied to calculate average results, and the output of the MS-SSIM quality metric is between 0 and 1. The higher value of the quality metric always represents that the actual image is closer to decoded image.

\textbf{Visualization.} To verify the quality of the image, we visualize the performance of our reconstructed images in each iteration with our variable-rate model. The comparison results of our proposed network and other image codecs are presented in Fig. \ref{fig:Fig_4} respectively. The results are compared with the original image, George's ~\cite{002} (4:2:0). As illustrated in Fig. \ref{fig:Fig_4}, the proposed network with blocks gives a high similarity compare to the previous methods ~\cite{002}. The visual performance of JPEG and JPEG2000 has worst because the edges contain ringing artifacts. The results of HEVC based BPG are clearer, and smoother compare with JPEG. Compare with George's ~\cite{002} and JPEG in terms of SSIM. The PSNR score is a little lower than BPG and WebP (in sampling factor 4:2:0 format). However, the proposed variable-rate method achieves better MS-SSIM, especially at the tenth epoch.
\section{Conclusion}
In this paper, two effective blocks i.e., analysis and synthesis blocks, are proposed based on convolutional layers and GDN layers embedded into RNN based image compression network. The pixel RNN approach is adopted and constructs a pixel-wise binary quantization scheme with some hidden values using linear convolution. Furthermore, to more enhance the performance of the network, RNN cells are utilized. These cells are placed in the encoder and decoder sides to increase performance. The results illustrate that the framework with novel blocks outperforms George ~\cite{002} method and standard images such as BPG and JPEG.

\section{Acknowledgement}
This work was supported by Basic Science Research Program through the National Research Foundation of Korea (NRF) funded by the Ministry of Education (2020R1A6A1A03038540) and by Institute of Information \& communications Technology Planning \& Evaluation (IITP) grant funded by the Korea government (MSIT) (2019-0-00136, Development of AI-Convergence Technologies for Smart City Industry Productivity Innovation).

{\small
\bibliographystyle{ieee}
\bibliography{egbib}
}

\end{document}